\documentclass[times,twocolumn,final]{elsarticle}

\usepackage{subfigure}
\usepackage{medima}
\usepackage{framed,multirow}

\usepackage{amssymb}
\usepackage{latexsym}

\usepackage{url}
\usepackage{xcolor}

\usepackage{hyperref}

\definecolor{newcolor}{rgb}{.8,.349,.1}

\journal{Medical Image Analysis}
\graphicspath{{pics/}{figs/}}
\begin{document}

\verso{H. Hashemi \textit{et~al.}}

\begin{frontmatter}

\title{Ultrafast Ultrasound Imaging for 3D Shear Wave Absolute Vibro-Elastography}%

\author[1]{Hoda S. \snm{Hashemi}\corref{cor1}}
\cortext[cor1]{Corresponding author.}
\ead{hoda@ece.ubc.ca}
\author[1,2]{Reza \snm{Zahiri Azar}}
\author[1,3]{Septimiu E. \snm{Salcudean}}
\author[1,4]{Robert N. \snm{Rohling}}
%% Third author's email

\address[1]{Department of Electrical and Computer Engineering, University of British Columbia, Vancouver, BC, Canada.}
\address[2]{DarkVision Technologies Inc., Vancouver, BC, Canada.}
\address[3]{Department of Biomedical Engineering, University of British Columbia, Vancouver, BC, Canada.}
\address[4]{Department of Mechanical Engineering, University of British Columbia, Vancouver, BC, Canada.}

\received{13 September 2022}
\finalform{}
\accepted{}
\availableonline{}
\communicated{}

\begin{abstract}
%%%
Shear wave absolute vibro-elastography (S-WAVE) is a \textcolor{black}{3D quantitative} imaging technique that generates steady-state shear waves inside the tissue using multi-frequency excitation from an external vibration source.
In this work, plane wave imaging is introduced to reduce total acquisition time while retaining the benefit of \textcolor{black}{a} 3D formulation. Plane wave imaging with a frame rate of 3000 frames/s is followed by 3D absolute elasticity estimation. \textcolor{black}{This paper presents} two imaging sequences of ultrafast S-WAVE for \textcolor{black}{low frequency and high frequency} sets of excitation frequencies using a Verasonics system and a motorized swept ultrasound transducer \textcolor{black}{(3D wobbler)} to synchronize ultrasound acquisition with the external mechanical excitation. The overall data collection time is improved by 83-88\% compared to the original 3D S-WAVE because of the per-channel acquisition offered by the Verasonics system. \textcolor{black}{The method is validated on four} liver fibrosis tissue-mimicking phantoms and \textcolor{black}{initial results are shown} on \textit{ex vivo} bovine liver. The curl operator was previously used in magnetic resonance elastography (MRE) to cancel out the effect of the compressional waves. In this work, we apply the curl operator to the full 3D displacement field followed by 3D elasticity reconstruction. \textcolor{black}{The results of the phantom experiment show that by using the curl of a 3D displacement field, the accuracy of the elasticity estimation improves by 14\% with a decrease of the standard deviation (STD) of 18\%} compared to reconstruction using the curl of a 2D displacement field. The phantom results also demonstrate {12\% enhancement in accuracy of elasticity estimation and} 45\% \textcolor{black}{lower} STD \textcolor{black}{compared to the elasticity reconstruction} without the curl \textcolor{black}{on a 3D displacement field}. We also \textcolor{black}{show the} experimental results \textcolor{black}{of} a \textcolor{black}{standard} method based on acoustic radiation force impulse (ARFI) \textcolor{black}{on the phantoms and an \textit{ex vivo} sample}.
%%%%
\end{abstract}

\begin{keyword}
%% Keywords
\KWD \\
Ultrafast elastography\\ Shear wave absolute vibro-elastography\\ S-WAVE\\ 3D elastography\\ Ultrasound\\ Elasticity estimation
\end{keyword}

\end{frontmatter}

%\linenumbers

\section{Introduction}
\label{intro}
Measurements of low frequency mechanical vibrations inside the tissue such as shear waves contain useful information about tissue properties including elasticity. The conventional single-line delay and sum algorithm for ultrasound beamforming provides an imaging frame rate of 25 to 50 frames/s for an image with the depth of 5 cm and 128 lines. However, an acquisition frame rate higher than 1000 frames/s is needed for real-time tracking of shear waves which typically travel at speed of 1 to 10 m/s inside the tissue~\cite{tanter2014ultrafast}. Plane waves can be used to image the tissue at a frame rate of several {thousand frames/s} by parallel beamforming for all positions in the medium~\cite{tanter2002constrast, sandrin2002shear}. High frame rate imaging {can therefore} reduce \textcolor{black}{imaging} time.
Shear wave propagation should also be separated from the motion artifacts caused by breathing \textcolor{black}{in many patient examinations} as they take place on a slower time scale~\cite{tanter2014ultrafast}. Furthermore, local estimation of the tissue stiffness can be improved by higher spatial and temporal sampling rates~\cite{tanter2014ultrafast}.

Several elasticity imaging techniques have employed ultrafast elastography to increase the frame rate as well as the accuracy of the shear wave tracking. Some of these techniques use acoustic radiation force impulse (ARFI) to generate tissue motion from focused ultrasound beams. Multiple Track Location Shear Wave Elasticity Imaging (MTL-SWEI) includes methods such as Supersonic Shear waves Imaging (SSI)~\cite{bercoff2004supersonic}, and Comb-push Ultrasound Shear Elastography (CUSE)~\cite{song2012comb}, where a single ARFI excitation is generated inside the tissue and the shear waves are tracked in multiple locations. In this study, we combine ultrafast imaging with the 3D Shear Wave Absolute Vibro-Elastography (S-WAVE) method~\cite{abeysekera2015vibro} \textcolor{black}{to provide fast 3D data acquisition}, and call our method {Ultrafast S-WAVE}. In S-WAVE, similar to magnetic resonance elastography (MRE), an external mechanical excitation source generates steady-state shear waves at multiple low frequencies throughout an entire {volume}. By estimating the tissue displacements, the wave equation can be used to estimate the shear modulus ($\mu$). By modeling the tissue as a linear, elastic and nearly incompressible material, the elasticity (stiffness) is approximated using the shear modulus through $E = 3 \mu$.

Elastography is beneficial in detection, staging and monitoring of the liver disease\cite{tapper2018noninvasive}. Nonalcoholic fatty liver disease (NAFLD) is the {most} common cause of chronic liver disease which has affected more than 25$\%$ of the global population~\cite{loomba2013global} especially in individuals with obesity, hypertension, and diabetes. The progressive version of NAFLD can lead to the fibrosis, cirrhosis, and finally liver death. 
Furthermore, it has been found that the stage of fibrosis is exponentially linked to the mortality rate of NAFLD~\cite{dulai2017increased}. Therefore, determination of the liver fibrosis is important in patient treatment and can be diagnosed using biopsy~\cite{de2016diagnose} or {liver} imaging evidence. 

%%%%%%%%%%%%%%%%%%%%%%%%% Fig  : experiment setup
\begin{figure*}[!t]
	\begin{center}
		\includegraphics[width=.80\textwidth] {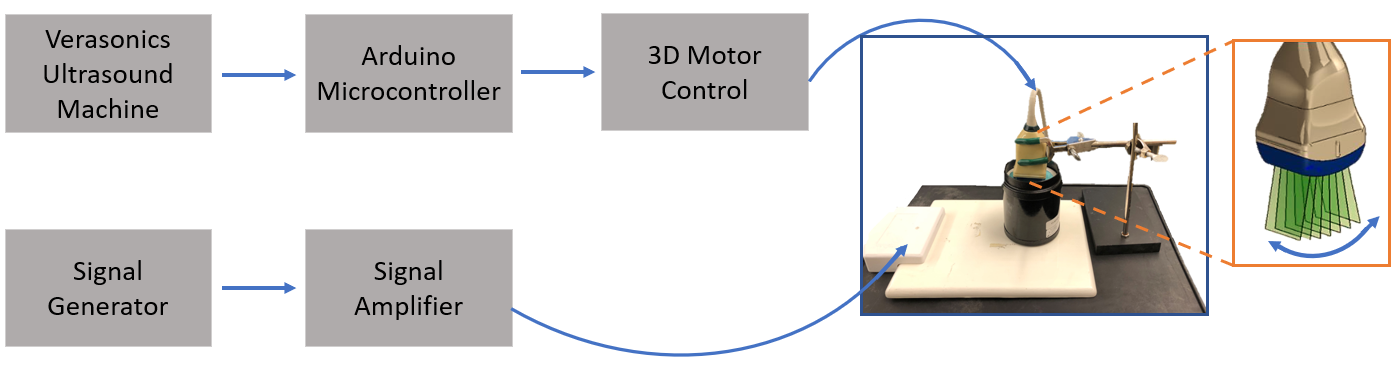}
		\caption{\textcolor{black}{Ultrafast S-WAVE setup. The Verasonics ultrasound machine sends signals to the Arduino microcontroller to sweep the imaging plane of the wobbler transducer (4DL14-5/38) through a motor controller box. A signal generator and amplifier are connected to a board excitation source which generates shear waves inside the tissue. The transducer picture is magnified to show the imaging planes and the transducer sweep direction.}}
		\label{fig:experimentSetup}
	\end{center}
\end{figure*}

Imaging, as a non-invasive approach, is preferred over liver biopsy, however liver fibrosis has no \textcolor{black}{direct} signature detectable by current imaging techniques. Therefore, all the imaging modalities attempt to detect fibrosis indirectly. Stiffness is the leading imaging biomarker for detecting and staging liver fibrosis~\cite{dulai2016mri} {which} can be accessible through elastography.
Some of the {commercial} elastography methods that are clinically used to estimate liver stiffness include: VCTE (Echosens), SWE (Hologic), MRE, ARFI (Philips, Siemens), and S-WAVE (Sonic Incytes). MRE has greater accuracy compared to the other methods~{\cite{yin2018ultrasound}}, however it is more expensive and time consuming. S-WAVE is the most similar elastography method to MRE in terms of using an external excitation source to generate steady-state shear waves inside the liver at frequencies \textcolor{black}{in the range of} 40 to 70 Hz. The advantage of the low-frequency vibration source is the high penetration of the shear waves into the entire liver. S-WAVE has been successfully used in liver disease studies~\cite{zeng2020three} as well as other organs such as the placenta~\cite{abeysekera2017swave}, the breast~\cite{eskandari2013identifying,shao2021breast}, the prostate~\cite{moradi2014multiparametric}, and the kidney~\cite{schneider2016blood}. \textcolor{black}{Some of the previous studies reported limitations in using ARFI shear wave imaging for obese patients. In~\cite{cassinotto2013liver,sirli2013liver}, SSI failed in providing reliable estimations for 15\% of the subject and 49\% of the patients, respectively, for cases with Body Mass Index (BMI) greater than 30. A recent study~\cite{ornelas2020s3264} examined the accuracy of the VCTE known as Fibroscan in the diagnosis of liver fibrosis in clinical practice and concluded that BMI should be taken into account while selecting patients for the staging of fibrosis to alleviate inherent limitations. Given the fact that patients with nonalcoholic steatohepatitis (NASH) disease are commonly overweight individuals, the need for methods such as S-WAVE to work reliably for obese patients is clearly vital. In 3D S-WAVE~\cite{zeng2020three} and several clinical imaging methods, patients have to perform a breath hold of approximately 10 seconds~\cite{barr2015elastography} which is not always possible, especially for elderly people, postoperative patients, or paediatric population~\cite{riccabona2003potential}. Arthritis in the neck or shoulders~\cite{dershaw2000imaging}, obesity or cardiopulmonary disease~\cite{reiner2013contrast} can also cause difficulties in holding breath. Therefore, faster imaging techniques are needed to decrease the ultrasound exam time for patients.}

\textcolor{black}{Standard 2D shear wave elastography methods implemented on clinical scanners showed good performance in assessing liver fibrosis but unsatisfying results in assessing the entire spectrum of steatohepatitis (fibrosis, steatosis, inflammation, ballooning) leading to cirrhosis and liver cancer.} 3D ultrasound can provide the motion measurement in all three directions over a volumetric region of interest which is the best approach for accurately estimating the shear modulus from the wave equation in elastography systems~\cite{eskandari2008viscoelastic}.
%Most of commercially available ultrasound elastography and MR elastography implementations sample wave propagation in only one or two dimensions. 
\textcolor{black}{Furthermore, the 3D implementation permits assessment of a larger liver volume and enables analysis of a larger number of tissue slices. It provides improvements in colocalization across time points, monitoring for longitudinal change, and a more complete representation of wave propagation~\cite{tang2015ultrasoundP2}.}

In this work, we use a \textcolor{black}{3D} motorized array transducer {to acquire volumetric ultrasound data} and present imaging sequences and data processing using \textcolor{black}{the} Verasonics \textcolor{black}{research} system. The excitation and imaging sweep are synchronized to enable {quasi-instantaneous} 3D imaging which is followed by the 3D motion estimation and elasticity reconstruction over a volume. Measuring motion in three directions has several advantages such as enhancements in accuracy of the shear wave motion estimation and speed measurements~\cite{wang2013precision},  {opportunity to} incorporate anisotropic properties of the tissue~\cite{wang2013imaging}, and decreasing diffraction bias in shear wave fields~\cite{yin2008diffraction}. 
% \textcolor{black}{Considering anisotropic feature of soft tissue is beneficial in grading liver fibrosis that cannot be measured using the Fibroscan device. The reason lies in the constitution of hepatic fibrosis that can be considered to be homogeneous from a macroscopic point of view, but is highly heterogeneous and anisotropic on the microscopic level~\cite{sandrin2003transient}.}
% Viscous loss, anisotropy, heterogeneity and nonlinearity are particularly interesting properties that cannot be measured using the present form of the Fibroscan®.

Utilizing the curl of the displacement field in elasticity reconstruction, which needs 2D or 3D motion estimation, eliminates the contribution of the compressional waves~\cite{baghani2009theoretical}. It was previously used in MRE where measurements of \textcolor{black}{all} three motion components are available~\cite{glaser2012review,manduca2018waveguide}. In this work, we use the curl of the 3D displacement phasors in 3D elasticity reconstruction. To the best of our knowledge, it is the first instance that the curl of a 3D displacement map obtained from a single ultrasound transducer is used in solving the inverse problem of elastography to calculate the tissue elasticity map. In this paper, two imaging sequences in a Verasonics programming scripts with different sets of excitation frequencies are described where the overall data collection time is equal to 1.5 and 2 s for each imaging sequence. \textcolor{black}{The challenge with defining a new sequence is that the imaging time should be selected such that the acquisition is synchronized with the mechanical excitation. In other words, the overall period of multi-frequency signal determines the imaging time which is discussed in \ref{sync} in details.} The high frame rate of 3000 frames/s, and deep penetration of low excitation frequencies has a goal of providing elasticity measurements at both shallower and greater depths. The method is \textcolor{black}{validated} on phantoms and \textcolor{black}{also applied on} an \textit{ex vivo} bovine liver \textcolor{black}{to show some initial results on real tissue}. The results of the proposed method are compared with an implementation of the MTL-SWEI technique on the \textcolor{black}{same} Verasonics system~\cite{deng2016ultrasonic}. %\textcolor{black}{We believe the designed imaging sequences and microcontroller programs are beneficial to researchers interested in 3D ultrafast imaging with the Verasonics system and wobbler transducers and can be found at https://people.ece.ubc.ca/hoda/.} 
\textcolor{black}{\textit{In vivo} data will be acquired in future studies to detect and stage liver fibrosis in patients using the proposed method.} 

%When  this  script  is  executed,  the  Verasonics  system  entersreal-time   B-mode   imaging   mode,   with   a   B-mode   imageshown  on  the  computer  screen.  When  the  user  clicks  theB-mode image on  the screen, the sequence then enters shearwave  imaging  mode and  executes the  MTL-SWEI sequence. We will also refer to this C5-2 example script in the followingsections.

\textcolor{black}{In a previous study on 3D S-WAVE published by our group~\cite{zeng2020three}, a matrix array transducer with a Philips ultrasound machine was used to collect sector-based volumetric data with an effective frame rate of 250 Hz and axial displacement measurements over a volume. The advantages of this work over the previous work are as follows:
	\begin{enumerate}
		\item Integrating a high frame rate imaging approach with the frame rate of 3000 frames/s to the S-WAVE system to improve the overall exam time from 12s to less than 2s. 
		\item Eliminating potential sector artifacts by using plane wave imaging that fires all transducer elements at the same time. 
		\item Using full 3D displacements over the volume which is more accurate compared to the case where only the axial measurement is used over the volume~\cite{hashemi20203d}.
		\item Utilizing the curl of the 3D displacements to ignore the effect of compressional waves.
	\end{enumerate}}
\textcolor{black}{Since the element to channel mapping for our current Philips matrix array is not compatible with the Verasonics ultrasound system, we used a 3D wobbler transducer in this work.}
% The implemented codes are publicly available from the Web site of the corresponding author.
The rest of this paper \textcolor{black}{is} organized as follows: Section~\ref{methods} details the imaging setup, parameter selection for the experiments, controlling the transducer motor and synchronizing the imaging sweep with the excitation source. Section~\ref{results} demonstrates the results of the proposed algorithm on phantoms, and \textit{ex vivo} data, and compares the elasticity values with a previous method using MTL-SWEI technique~\cite{deng2016ultrasonic} where plane waves are also utilized. \textcolor{black}{The goal is to investigate the consistency of the elasticity ranges especially in \textit{ex vivo} sample achieved by both methods.} The results of the reconstruction with and without the curl of displacement field are demonstrated. Section~\ref{discussion} discusses some of the limitations in experiments and possible future work. Finally, our conclusions are presented in Section~\ref{conclusion}.

	 %% The differences are as follows. First, the introduction and the literature review of this paper are more comprehensive. Second, this paper includes the results of finite element and Field II simulations as well as data from seven patients, whereas the conference paper did not include simulation results and was further limited to two patients only. Finally, the results are discussed in substantially more detail in this extended paper.}

% More specifically, the contributions of this work are as follows.
%introducing a novel method based on ultrasound elastography for comparing affected and unaffected arms in patients with lymphedema;
%proposing a novel method for TDE of quasi-static elastography that works reliably with the challenging patient data.

%%%%%%%%%%%%%%%%%%%%%%%%%%% Methods 

\section{Methods}
\label{methods}

\subsection{Proposed Technique: UltraFast Shear Wave Absolute Vibro-Elastography (Ultrafast S-WAVE)}
\label{method_UF_SWAVE}

In S-WAVE~\cite{abeysekera2015vibro}, an external excitation source is \textcolor{black}{placed against or beneath} the tissue sample or patient \textcolor{black}{as shown in Figure~\ref{fig:experimentSetup}.} The {signal generator} is then programmed to generate sinusoidal vibrations at multiple frequencies leading to steady-state shear waves inside the tissue. For low frequency sequences which can penetrate deeper inside the tissue {in} application{s} such as {the placenta~\cite{rac2015ultrasound}} and the liver{~\cite{zeng2020three}}, we use three excitation frequencies of 40, 50, and 60 Hz. {These frequencies} are \textcolor{black}{within} the typical range utilized by some of the previous methods such as MRE, transient elastography~\cite{barr2015elastography}, and crawling wave sonoelastography~\cite{ormachea2016shear} for imaging the liver. The higher frequency sequence using the excitation frequencies of 100, 160, and 200 Hz are designed for {shallower} depth applications such as the {breast}.
A Vantage 256 ultrasound system (Verasonics Inc, Kirkland, WA) equipped with a 4DL14-5/38 motorized swept volume \textcolor{black}{(3D wobbler)} transducer (Vermon, Tours, France) with {a} center frequency of 5.1 MHz and the sampling frequency of 20.4 MHz is used in all Ultrafast S-WAVE experiments. A custom motor controller, which has been used in previous studies~\cite{abeysekera2015vibro,abeysekera2017swave,deeba2021multiparametric}, is connected between the Vantage ultrasound machine and the motorized ultrasound transducer to \textcolor{black}{control the transducer sweep inside the housing and} acquire volumetric data. An Arduino microcontroller is also used to generate input signals for the motor \textcolor{black}{controller} {to synchronize data acquisition between motor movements}. \textcolor{black}{As shown in Figure~\ref{fig:experimentSetup}, inside the transducer housing, a linear array transducer sweeps back and forth generating 2D ultrasound images which are used to reconstruct the 3D ultrasound volume. In this work, the transducer acquires 10 planes and collects several frames at each plane location. The frames are grouped together such that the slices within a volume are acquired at a common excitation phase.} We use {an imaging depth of 5 cm for the phantom elasticity measurements. \textcolor{black}{To be consistent} \textcolor{black}{with} the \textit{ex vivo} experiment, an imaging depth of 5 cm is used for \textcolor{black}{both} lower and \textcolor{black}{higher} frequency sequences \textcolor{black}{while} all the \textcolor{black}{reported} measurements are performed at the depth of 3 cm for the \textit{ex vivo} sample} \textcolor{black}{with} a frame rate of 3000 frames/s to fully capture the shear wave propagation and tissue displacement. 
The radio-frequency (RF) data is stored and \textcolor{black}{beamformed~\cite{garcia2021make,perrot2021so} in a rectangular grid for each 2D frame. The corressponding frames for each volume are placed together and form a volume.} Axial, lateral, and elevational displacement maps are calculated using the GLUE3D algorithm~\cite{hashemi20203d} which has been used previously in different elastography methods~\cite{hashemi2018assessment,hashemi20203d,hashemi2017global}. \textcolor{black}{Similar to the other elastography methods, it has tunable parameters ($\alpha_{1,2}$, $\beta_{1,2}$, $\gamma_{1,2}$, and $t$) which regulate the spatial and temporal displacement continuity throughout the data volume. The parameters can be set at the beginning of the ultrasound exam based on the target organ. Higher coefficient values lead to smoother displacement maps. In all experiments, we set the tunable parameters of the GLUE3D algorithm to the recommended values in~\cite{hashemi20203d} such that $\alpha_{1}=\alpha_{2}=\beta_{1}=\beta_{2}=\gamma_{1} = 200$, $\gamma_{2}= 0.5$, and $t=0.001$. Changing the GLUE3D coefficients by 100$\%$ changes the elasticity values by less than 2$\%$ for both phantom and \textcolor{black}{\textit{ex vivo}} data.} 
Displacement estimation is followed by fitting displacement phasors~\cite{abeysekera2016three}, and calculating the curl of 3D displacement phasors to be utilized in tissue elasticity measurement using local frequency estimation (LFE)~\cite{oliphant2000error}.

\begin{figure}[tb]
	\begin{center}
		\subfigure[]{\label{fig:1a}\includegraphics[width=.48\textwidth] {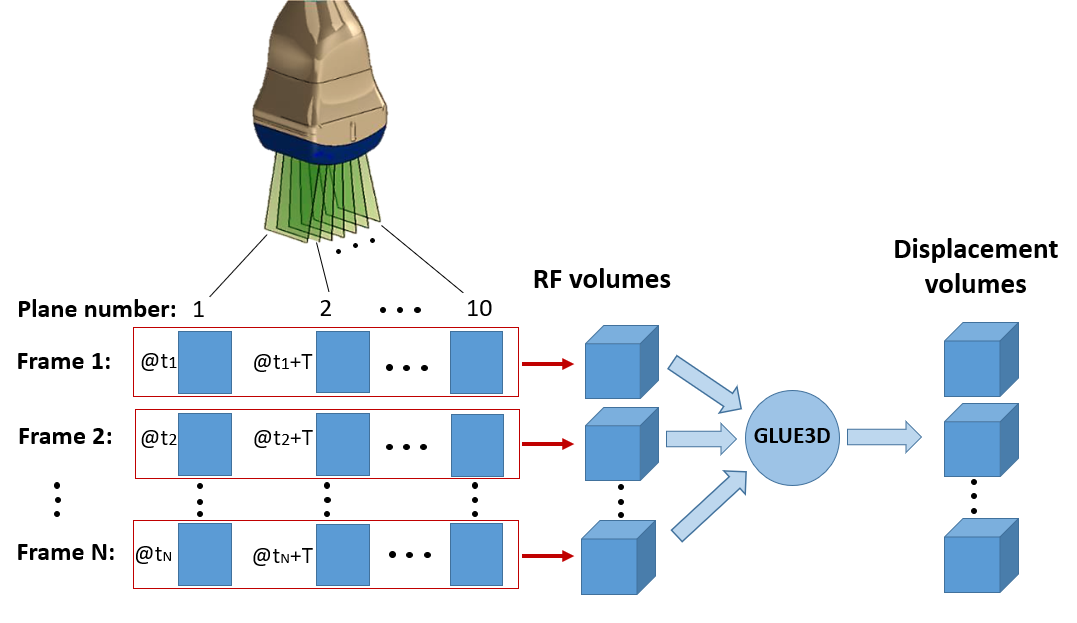}}\\
		\subfigure[]{\label{fig:1a}\includegraphics[width=.48\textwidth] {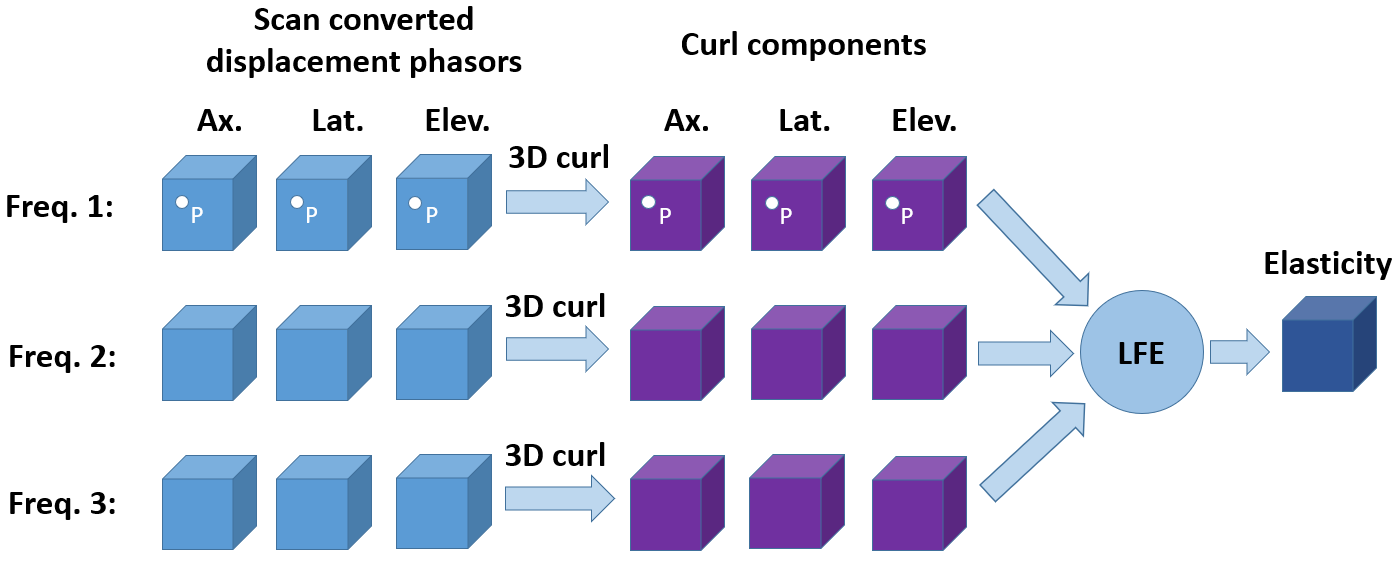}}
		\caption{Elasticity estimation pipeline. (a) shows RF volume \textcolor{black}{acquisition} and displacement estimation. Size of the RF volumes are shown in samples. (b) demonstrates the process of applying the curl operator on displacement phasors. For each excitation frequency, there are axial, lateral, and elevational phasors. Performing 3D curl for a point P in the volume results in a curl vector with three components. Applying the 3D curl on all of the points within the volume provides three volumes as axial, lateral and elevational curl components which are used in LFE function to produce the elasticity volume.}
		\label{fig:curl}
	\end{center}
\end{figure}

Figure~\ref{fig:curl} (a) shows the formation of $N$ RF volumes. The planes inside each RF volume are acquired at a common excitation phase given {the} periodic motion of the excitation wave. The transducer {volume} sweeps {comprise} 10 planes \textcolor{black}{(slices)}. $N$ frames {are acquired at} each plane location such that the frames for the first plane are collected at $\{t_{1}, t_{2}, ..., t_{N}\}$, for the second plane at $\{t_{1}+T, t_{2}+T,..., t_{N}+T\}$, and so forth where T is the period of the excitation signal and $t_{2}-t_{1}=1/(Frame Rate)$. The planes acquired at $\{t_{1}, t_{1}+T, t_{1}+2T, ...\}$ are then grouped into the first volume, the planes collected at $\{t_{2}, t_{2}+T, t_{2}+2T, ...\}$ \textcolor{black}{are grouped in} the second volume, and etc. The \textcolor{black}{spacing of the beamformed RF data and displacement volumes} in the axial, lateral and elevational directions are 0.15 mm, 0.48 mm, and 0.45$^{\circ}$, respectively. The consecutive volumes are used by GLUE3D for 3D motion estimation and a sequence of displacement volumes $f(x,y,z,t)$ is generated. \textcolor{black}{Displacement phasors $F(x,y,z,w)$ are obtained at each spatial location and each frequency $w$ by taking the Fourier transform of the displacements.} The phasors are then scan converted to a 3D Cartesian grid with the uniform sample spacing of 0.5 mm \textcolor{black}{in all three directions}. \textcolor{black}{These} are demonstrated in \textcolor{black}{Figure~\ref{fig:curl}} (b). For each excitation frequency, there are three phasor volumes: axial, lateral, and elevational displacements, which are used in the 3D curl calculations.
{The} Curl of a 3D volume can be estimated as follows:\\
$\nabla\times \textbf{F} = (\frac{\partial F_{z}}{\partial y}-\frac{\partial F_{y}}{\partial z}) \hat{x} + (\frac{\partial F_{x}}{\partial z}-\frac{\partial F_{z}}{\partial x}) \hat{y} + (\frac{\partial F_{y}}{\partial x}-\frac{\partial F_{x}}{\partial y}) \hat{z} $,
where F is the 3D displacement phasor, x, y and z are axial, lateral, and elevational directions, respectively. A 3D grid of points is defined based on the uniform sample spacing in each direction. The partial derivatives in \textcolor{black}{the} curl formula are \textcolor{black}{calculated} using the central difference method. Note that applying the curl operator for a specific point P in the volume results in a 3D vector. Therefore, by performing the curl operator for all the points within the volume, three volumes corresponding to axial, lateral and elevational directions will be achieved and used in 3D LFE function.
To show the advantage of reconstruction using the 3D curl of displacements, the elasticity {of the phantoms are} estimated using two other approaches: first, reconstruction with no curl (using axial, lateral, and elevational displacement phasors); Second, reconstruction using 2D curl (curl of axial and lateral displacements i.e. the $\hat{z}$ {or elevational} component of $\nabla\times \textbf{F}$). 
%%%%%%%%%%%%%%%%%%%%%%%%% Fig  : curl calculation

%%%%%%%%%%%%%%%%%%%%%%%%% Fig  : Imaging sequences
\begin{figure*}[tb]
	\begin{center}
		\subfigure{\includegraphics[width=.85\textwidth] {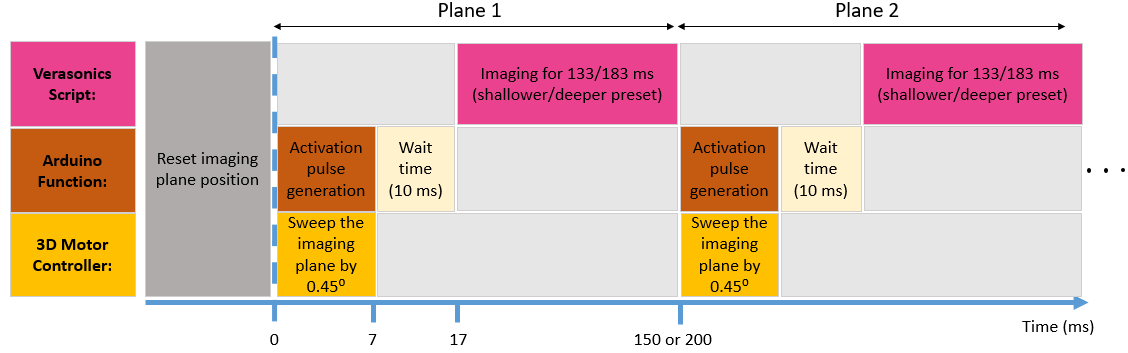}}\\
		\caption{\textcolor{black}{The imaging sequence used in the Verasonics script for the Ultrafast S-WAVE method. First, the imaging plane is reset to an initial plane position. The sequence starts by generating four activation pulses using an Arduino microcontroller to mechanically sweep the imaging plane inside the transducer housing by 4 steps each time to get an angle of $0.45^{\circ}$ between slices within the data volume, following by 10 ms for the motor settling time and 133 ms for imaging using shallower presets and 183 ms for deeper presets. This process is repeated for each plane.} }
		\label{fig:ImgSeq}
	\end{center}
\end{figure*}

\subsubsection{Ultrafast S-WAVE Imaging Sequence for Verasonics \& Parameter Specification}
\label{parameter_selection}

Parameter selection for S-WAVE sequences and guidelines of synchronizing excitation source, imaging sequence and the stepper motor inside of the transducer are explained in this section. We design two different sequences that can be used for imaging at \textit{``deeper'' depth} inside the organs such as human liver and placenta and also for \textit{``shallower''} imaging applications such as the breast.

\textcolor{black}{The wobbler transducer is defined in \textcolor{black}{the} Verasonics script as a custom transducer based on the transducer information and geometry. When our custom Verasonics script is executed, the Arduino microcontroller function resets the array location inside the transducer housing and moves it to an initial location adjusted for the start of imaging. Then the imaging starts and several RF frames are acquired.} \textcolor{black}{Figure~\ref{fig:ImgSeq} shows the sequencing used in the Verasonics script for the Ultrafast S-WAVE method. First, second, and third rows show the actions that is performed by the Verasonics machine, Arduino microcontroller and 3D motor controller while all the commands and functions are defined and called inside the Verasonics script. First, the imaging plane is reset to the middle plane of the transducer. Then, the sequence starts by generating four activation pulses using an Arduino microcontroller to mechanically sweep the imaging plane inside the transducer housing by $0.45^{\circ}$ followed by the motor settling time (10 ms). The Verasonics machine starts collecting data such that the number of acquired frames and imaging time depends on the \textit{shallower} or \textit{deeper} presets selected by the user at the beginning. When the imaging completes, the transducer sweeps to the next location while the imaging is paused in the script during the transducer sweep. This process is repeated until the last imaging plane. The imaging time for each preset depends on the overall period of the excitation signal. For the \textit{deeper} imaging preset, the excitation frequencies are 40, 50, and 60 Hz which result in an overall excitation period of 100 ms. The Vantage system is programmed to perform imaging for 183 ms acquiring 549 RF frames. Therefore, the overall time spent on each plane is 200 ms which is a factor of the excitation period to synchronize ultrasound imaging with mechanical excitation (details are described in ~\ref{sync}). By repeating the same sequence, the total number of 10 planes (slices) are imaged to form an RF volume of data resulting in the overall exam time of 2 s (5490 frames).} For the excitation frequencies of 100, 160, and 200 Hz, the overall excitation period is 50 ms for imaging at \textit{shallower} depth in the tissue. The imaging duration for each plane is 133 ms. Therefore, the overall ultrasound examination time for a volume consists of 10 planes is 1.5 s (3990 frames).

\subsubsection{Motor Controller}
\label{controller}

For the motorized 3D transducer, the imaging plane is swept over a volume using a customized motor control component to collect volumetric data. \textcolor{black}{\textcolor{black}{The} imaging script sends signals to \textcolor{black}{the} motor microcontroller. \textcolor{black}{The microcontroller} is programmed before the start of imaging such that during the data collection the function is called within the Verasonics script while there is no need to \textcolor{black}{repeatedly program each} frame acquisition, thereby avoiding any programming delay.} The direction of the sweep, the angle between the planes, and the sweep activation pulse can be adjusted using the \textcolor{black}{imaging script and} \textcolor{black}{the} microcontroller.

%%%%%%%%%%%%%%%%% Table 1
\begin{table*}[ht]
	\centering
	\caption{The average elasticity values $\pm$ standard deviations of the CIRS liver fibrosis phantoms in [kPa] {using the excitation frequencies of 40,50,60 Hz}. A window of size $1\times1\times1$ $cm^3$ at the depth of $5$ $cm$ is considered for elasticity calculation of mean$\pm$std.} 
	\label{tab:E_phantom}
	\begin{tabular}{|p{2.25cm}|p{2cm}|l|p{4cm}|p{3cm}|}%{c c c c c c} 
		\hline
		%\toprule
		Manufacturer  & ARFI &   Ultrafast S-WAVE  &   Ultrafast S-WAVE &   Ultrafast S-WAVE \\
		value &   &   (no curl) &   (2D curl) &   (3D curl)\\
		\hline
		2.8 & 2.32 $\pm$ {0.01}  & 2.03 $\pm$ {0.08} & 2.11 $\pm$ {0.03} & 2.45 $\pm$ {0.03} \\
		
		6.2 & 5.08 $\pm$ {0.16}  & 5.00 $\pm$ {0.16} & 4.62 $\pm$ {0.13} & 5.92 $\pm$ {0.12} \\
		
		11.6 & 10.07 $\pm$ {0.11}  &  10.26 $\pm$ {0.49}  & 9.66 $\pm$ {0.98}  & 11.19 $\pm$ {0.47} \\
		
		21.2 & 19.27 $\pm$ {0.13}  & 19.07 $\pm$ {0.27} & 18.96 $\pm$ {0.36} & 20.98 $\pm$ {0.20} \\
		% average of stds :0.25 0.375 0.205
		% imp in stds: 18% 45%
		%\toprule
		\hline
	\end{tabular}
\end{table*}

\subsubsection{Synchronization of excitation source, imaging, and transducer sweep} 
\label{sync}

\textcolor{black}{In this work, the wobbler transducer sweeps 10 slices and collects several frames at each slice location.} For \textcolor{black}{the} \textit{deeper} imaging setting (40, 50, and 60 Hz), the excitation signal period is 100 ms. In order to synchronize ultrasound acquisition with mechanical excitation, the overall data acquisition time for each plane should be a factor of the excitation signal period (i.e. 100 ms, 200 ms, etc). So, for each plane, the overall acquisition time was adjusted to 200 ms.
%So, for each plane, the overall acquisition time should be a factor of 100 ms (i.e. 100 ms, 200 ms, etc). Although 100 ms provides faster acquisition time, {evaluation of the Fourier spectrum of the displacement shows that this is not long enough} to distinguish the three excitation frequencies {of the displacement representation in {the} frequency domain}. To have three dominant peaks in {the} frequency spectrum {that} are within 5\% of the excitation frequency values, the overall acquisition time should be adjusted to at least 200 ms.
The synchronization between the imaging and transducer sweep is done within the Vantage sequencing script. The \textcolor{black}{microcontroller} function is called to generate \textcolor{black}{four step} pulses to \textcolor{black}{sweep} the transducer by 0.45$^{\circ}$ followed by a wait time of 10 ms. Afterward, the imaging process starts and continues for 183 ms. This process is repeated until the last imaging plane is acquired. For the \textit{shallower} setting (100, 160, and 200 Hz), the overall excitation period is 50 ms. So, \textcolor{black}{we used the acquisition time of 150 ms which is a factor of the excitation signal period. In both deeper and shallower settings, the excitation frequencies will be visible in the displacements in frequency domain.}

\subsection{Liver Fibrosis Phantom}
\label{method_phantom}

The RF data is acquired from four elastic liver fibrosis phantoms (Model 039, CIRS Inc., Norfolk, VA, USA) {with the imaging depth of 5 cm} using the Vantage 256 ultrasound system {and \textit{deeper} setting}.
According to the manufacturer, the elasticity value of the phantoms are 2.8, 6.2, 11.6, and 21.2 kPa with a precision of $\pm 4\%$. It has a sound speed of 1540 m/s, ultrasound attenuation of 0.5 dB/cm/MHz, and the overall depth {of} 10 cm. 

\subsection{\textit{Ex vivo} Bovine Liver}
\label{method_exvivo}
The \textit{ex vivo} data {was} collected from bovine liver tissue as shown in Figure~\ref{fig:experimentSetup}. The sample has a depth of $6$ cm in {the} axial direction. Both {the} ARFI and S-WAVE {acquisition} methods are performed at room temperature ($20$ $^{\circ}$C). {Imaging depth of 5 cm is used for both \textit{shallower} and \textit{deeper} settings \textcolor{black}{to have a fair comparison}. All the \textcolor{black}{reported} measurements are obtained at the depth of 3 cm.}

\subsection{ARFI: Multiple Track Location Shear Wave Elasticity Imaging (MTL-SWEI)}
\label{method_ARFI}
MTL-SWEI is a group of elasticity estimation techniques based on ARFI. A single push from the focused transducer beam is induced {into} the tissue to generate tissue motion. Multiple track locations are used to estimate the group shear wave speed (SWS) and phase velocity over the frequency content of the shear waves. We use an implementation of the MTL-SWEI on {the} Verasonics system where a single push with the frequency within the lower $-6 dB$ bandwidth of the transducer will be generated at the center of the transducer following by plane wave tracking, and IQ beamforming~\cite{deng2016ultrasonic}. {\textcolor{black}{T}here are zero delays across the curvilinear transducer elements resulting in diverging waves due to the curved transducer surface. We perform 8 measurements at a single location by slightly rotating the medium about that location between each acquisition. As such, we have a different speckle realization for the same area at each measurement.} The elasticity can be estimated by considering the tissue as an isotropic incompressible elastic material and using $E=3\times \rho \times {C_s}^2$, where $\rho$ is the tissue density which is considered $1000$ $\frac{kg}{m^3}$ for soft tissues and ${C_s}$ is the group SWS. For the rest of this paper, we use the general term ``ARFI'' to refer to this method. A C5-2v curvilinear transducer with the center frequency of 3.6 MHz and the sampling frequency of 14.24 MHz is used in all ARFI experiments.

%%%%%%%%%%%%%%%%%%%%%%%%% Fig  : phantom phasors
\begin{figure}[ht]
	\begin{center}
		\subfigure[Axial displacement phasor ]{\label{fig:1a}\includegraphics[width=.32\textwidth] {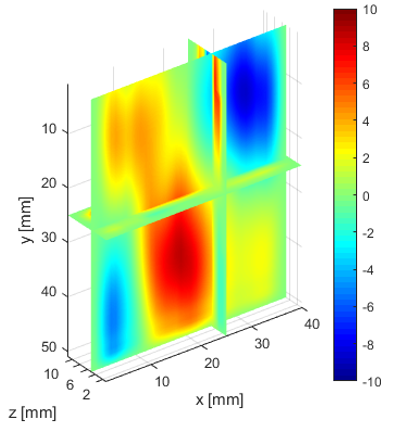}}
		\subfigure[Lateral displacement phasor]{\label{fig:1a}\includegraphics[width=.32\textwidth] {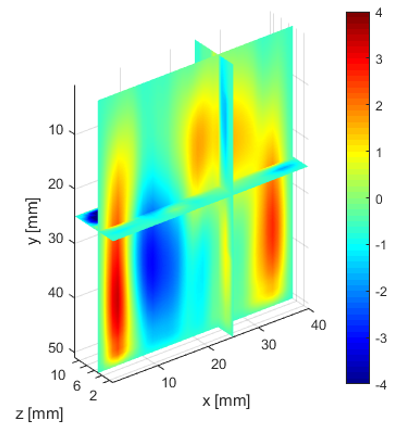}}
		\subfigure[Elevational displacement phasor]{\label{fig:1a}\includegraphics[width=.32\textwidth] {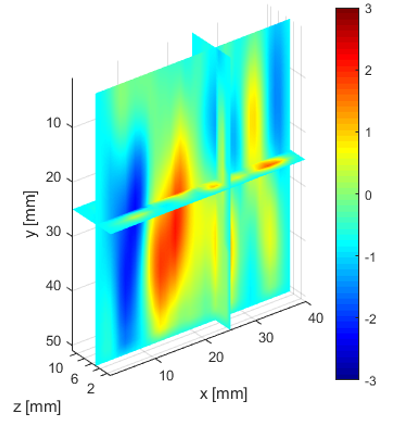}}	
		\caption{Estimated full 3D displacement phasors of the phantom data for the excitation frequency of 50 Hz. \textcolor{black}{Colour bar unit is $[\mu m]$ displacement.}}
		\label{fig:phantom_phasors}
	\end{center}
\end{figure}

%%%%%%%%%%%%%%%%%%%%%%%%% Fig  : exvivo freq sweep plot
\begin{figure}[tb]
	\begin{center}
		\subfigure[\textcolor{black}{ARFI}]{\label{fig:1a}\includegraphics[width=.4\textwidth] {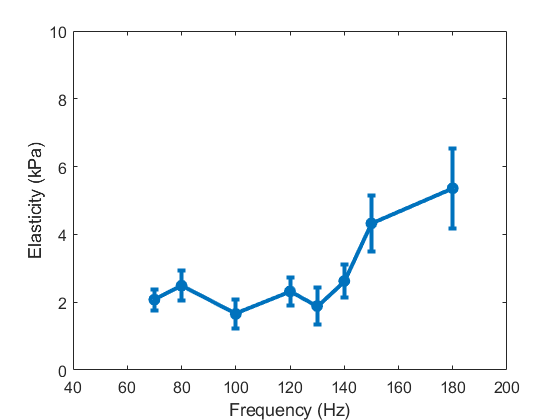}}
		\subfigure[Ultrafast S-WAVE]{\label{fig:1a}\includegraphics[width=.38\textwidth] {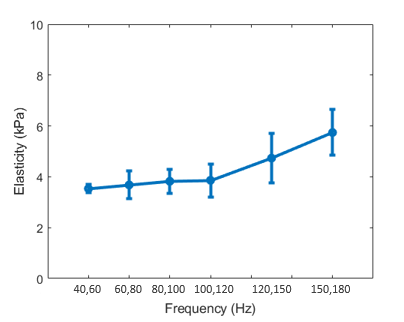}}
		\caption{Estimated elasticity values for a bovine liver sample based on the excitation frequencies. (a) and (b) show the estimated values using the ARFI and Ultrafast S-WAVE methods, respectively. In (b), each pair of excitation frequency is shown in horizontal axis, \textcolor{black}{and the elasticity measurments are averaged for each pair.}}
		\label{fig:exvivo_E_vs_freq}
	\end{center}
\end{figure}

%%%%%%%%%%%%%%%%%%%%%%%%%%%%%%%%%%%%%%%%%%%%%%%%%%%%%%%%%% Result %%%%%%%%%%%%%%%%%%%%%%%%%%%%%%%%%%%%%%%%%%%%%%
\section{Results}
\label{results}

\subsection{Phantom Results}
\label{results_phantom}

\subsubsection{ARFI}
RF data is first collected and analyzed using the ARFI method explained in~\ref{method_ARFI}. The average phase velocities for frequencies less than 200 Hz estimated by ARFI at the depth of $5$~cm {within} the phantom are used to calculate the elasticity values for all phantoms which are shown in Table~\ref{tab:E_phantom}.

\subsubsection{\textcolor{black}{Ultrafast} S-WAVE}
A set of 3D RF data is collected using the proposed Ultrafast S-WAVE technique described in~\ref{method_UF_SWAVE} and vibration frequencies of 40, 50, and 60 Hz. Figure~\ref{fig:phantom_phasors} shows the axial, lateral, and elevational phasors of a {slice of the} phantom for the excitation frequency of 50 Hz. The elasticity values for homogeneous phantoms are calculated using the curl of 3D displacement and LFE reconstruction. To be consistent with the ARFI measurements, a ROI of size $1\times1\times1$ $cm^3$ is considered at the same depth of $5$ cm. The average elasticity {with the standard deviation (STD) values} inside the ROI {are} shown in Table~\ref{tab:E_phantom}. Changing the size of the ROI by 100$\%$ will change the average elasticity value by less than 5$\%$.  
The elasticity volume is reconstructed using no curl, 2D curl and 3D curl. The results in Table~\ref{tab:E_phantom} confirm that using the curl of the 3D displacement map provides closer elasticity values to the manufacturer reported values. Furthermore, the elasticity values measured by the ARFI and Ultrafast S-WAVE methods are both in reasonable agreement with the manufacturer elasticity values. \textcolor{black}{The average error between the proposed method using curl and the reported manufacturer values is 5\% which is a reasonable error by considering the effect of aging on phantom materials and also the 4\% precision error in measurements of manufacturer elasticity values as mentioned in the phantom data sheet. The statistical analysis shows no significant differences between the manufacturer and estimated elasticity values by the Ultrafast S-WAVE using 3D curl in reconstruction.} Using the 3D curl helps to get more accurate mean elasticity values and reduces the STD within the homogeneous phantoms. The average STD for the Ultrafast S-WAVE using curl on 3D motion field improved by 18\% and 45\% compared to the no curl (but using elevational) and 2D curl (no elevational) approaches, respectively. 
\textcolor{black}{The difference error in mean elasticity value of Ultrafast S-WAVE from the manufacturer value improved by 12\% and 14\% (averaged for all phantoms) compared to the no curl and 2D curl approaches. Therefore, the 3D curl method shows better performance compared to the no curl and 2D curl techniques.}

%%%%%%%%%%%%%%%%%%%%%%%%% Fig  : exvivo phasors & E
\begin{figure*}[h]
	\begin{center}
		\subfigure[Axial displacement phasor ]{\label{fig:1a}\includegraphics[width=.33\textwidth] {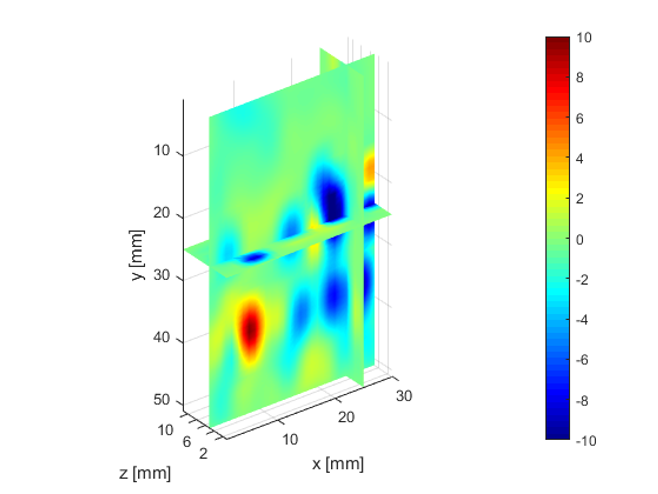}}
		\subfigure[Lateral displacement phasor]{\label{fig:1a}\includegraphics[width=.33\textwidth] {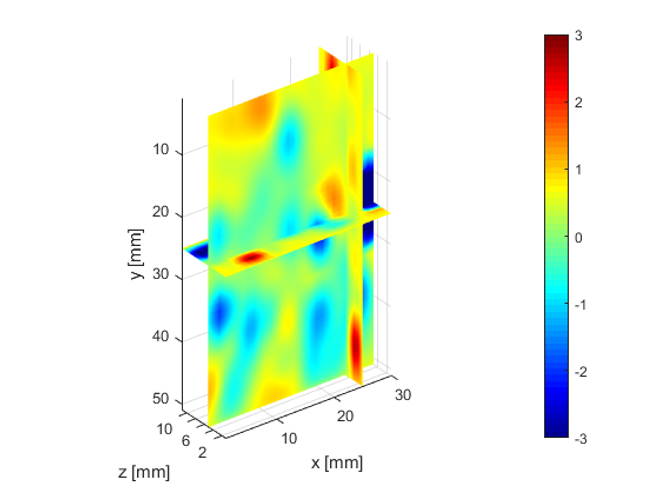}}
		\subfigure[Elevational displacement phasor]{\label{fig:1a}\includegraphics[width=.33\textwidth] {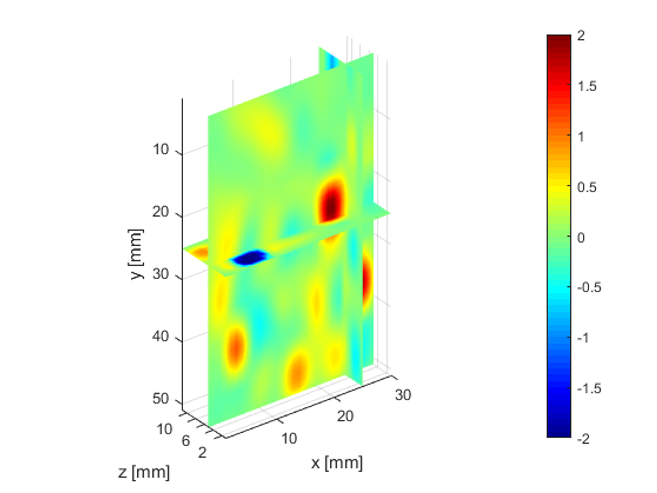}}\\
		\subfigure[B-mode]{\label{fig:1a}\includegraphics[width=.33\textwidth] {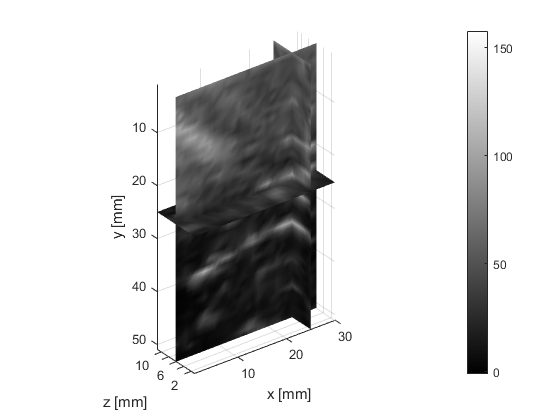}}
		\subfigure[Elasticity map ]{\label{fig:1a}\includegraphics[width=.33\textwidth] {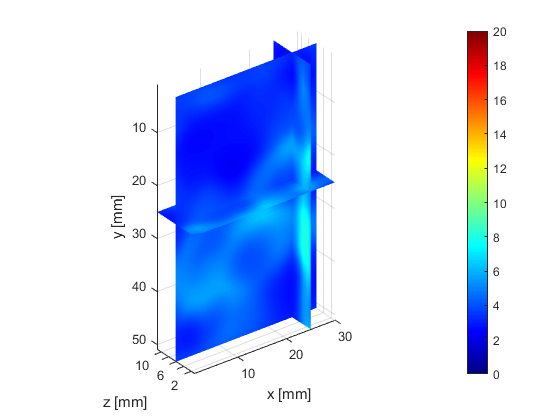}}
		\caption{Estimated displacement phasors, \textcolor{black}{B-mode} and elasticity map of the \textit{ex vivo} data. (a)-(c) are axial, lateral, and elevational phasor \textcolor{black}{volumes} from the Ultrafast S-WAVE method for excitation frequency of 200 Hz. \textcolor{black}{Colour bar unit is $[\mu m]$ displacement.} \textcolor{black}{(d) shows the B-mode volume.} The reconstructed elasticity \textcolor{black}{volume by averaging the elasticity maps of} all excitation frequencies (100, 160, 200 Hz) is shown in (e). \textcolor{black}{Colour bar unit is $[kPa]$ elasticity.}}
		\label{fig:exvivo_phasors}
	\end{center}
\end{figure*}

\subsection{\textit{Ex vivo} Results}
\label{results_exvivo}
The proposed two imaging sequences are applied on the \textit{ex vivo} bovine liver \textcolor{black}{to show some preliminary results on real tissue. As there is no ground truth for the actual elasticity of the \textit{ex vivo} sample, the result of the ARFI method is also provided to investigate the range of the elasticity values estimated by both methods}. All the \textcolor{black}{reported} measurements are performed at the depth of 3 cm which is half of the total depth of the tissue sample. 

\subsubsection{ARFI}
RF data are acquired as described in~\ref{method_ARFI}. Figure~\ref{fig:exvivo_E_vs_freq} (a) shows the elasticity values estimated using phase velocities for \textcolor{black}{different frequencies}. The elasticity values increase with frequency as the tissue mechanical characteristics of the liver have viscoelastic properties~\cite{palmeri2021radiological}. \textcolor{black}{For a} typical \textcolor{black}{low} frequency range of 70 to \textcolor{black}{180} Hz which \textcolor{black}{are among the frequencies} used by ARFI methods for the soft tissues~\cite{ormachea2016shear,song2015two} \textcolor{black}{and are close to the Ultrafast S-WAVE frequency range, the mean elasticity increases from 2.0 kPa to 5.4 kPa over the frequency range.} Due to the noise in phase velocity estimation for each frequency, some variances are visible among the generally increasing trend with frequency. Furthermore, some outliers in the time-of-flight trajectories can affect the results caused by the proximity to the capsule, vessels or poor coupling. \textcolor{black}{Higher frequencies typically show larger STD values since there is less energy at these frequencies. Furthermore, the waves are traveling faster at higher frequencies which may cause sampling issues. According to a previous study~\cite{wang2013precision}, the STD of time-of-flight will increase \textcolor{black}{with the square of the shear wave speed}. Therefore, smaller STD values are expected at lower frequencies.}
\textcolor{black}{As different elastography methods shows different values for the same sample/patient in ex/in vivo experiments~\cite{leung2013quantitative,potthoff2013influence}, the ARFI result is shown in this section to only investigate the dependency of the elasticity value to the frequency and also confirm the range of values provided by the proposed Ultrafast S-WAVE method described in \ref{results_Ultrafast_S-WAVE}.}
	
	%Although we expect a monotonic curve for the STD values, two points at the frequencies of 120 and 180 Hz show higher STD values compared to the neighboring frequencies causing by a few outliers in the time-of-flight trajectories for the phase velocity of these frequencies possibly due to proximity to the capsule or poor coupling of the gel and liver tissue. In general, phase velocities tend to be a much noisier measurement compared to group velocities because the energy is distributed along different frequencies.

\subsubsection{Ultrafast S-WAVE}
\label{results_Ultrafast_S-WAVE}
%To investigate the frequency dependency of the liver to the excitation frequency, multiple sets of data are acquired by increasing the vibration frequencies. Each time two frequencies are applied to the tissue. 

Before applying the main multi-frequency sequences described in \ref{parameter_selection}, different excitation frequencies in the range of 40 to 180 Hz are applied to the tissue to investigate the dependency of the elasticity on frequency for biological tissue. The estimated elasticity values for each pair of excitation frequencies are plotted in Figure~\ref{fig:exvivo_E_vs_freq}(b) showing that the average elasticity value increases with excitation frequency as we expect from the viscoelastic liver tissue~\cite{palmeri2021radiological}. \textcolor{black}{As different transducers and slightly different imaging location on the sample were used for the ARFI and Ultrafast S-WAVE experiments, direct comparison between values of both methods are not possible through this experiment.}

The \textit{deeper} and \textit{shallower} sequences described in~\ref{method_UF_SWAVE} are applied to the tissue \textcolor{black}{for the same imaging depth of 5 cm as the total sample depth is 6 cm, thereby comparing the results of two sequences at the same depth}. The 3D displacements are estimated using the GLUE3D method and the least square sinusoidal fitting. The elasticity values of 3.99$\pm$0.12 and 4.86$\pm$0.59 kPa are estimated respectively, using the LFE function on the curl of 3D displacement for two sets of excitation frequencies. \textcolor{black}{Note that for each frequency group, the reported elasticity value is the average of elasticity maps at each excitation frequency.} The values are similar to the reported values for animal livers~\cite{barry2014mouse,barry2015shear}. Figure~\ref{fig:exvivo_phasors} (a)-(c) shows the axial, lateral, and elevational phasors for the excitation frequency of 200 Hz. To avoid boundary artifacts, right and left boundaries of the RF frames are cropped by 0.7 cm. The elasticity map obtained \textcolor{black}{by averaging the elasticity maps of all} the excitation frequencies (100, 160 and 200 Hz) is depicted in (e). A ROI of size $1\times1\times1$ $cm^3$ at the depth of 3 cm is used for elasticity calculations. 

\begin{figure}[tb]
	\begin{center}
		\subfigure{\includegraphics[width=.50\textwidth] {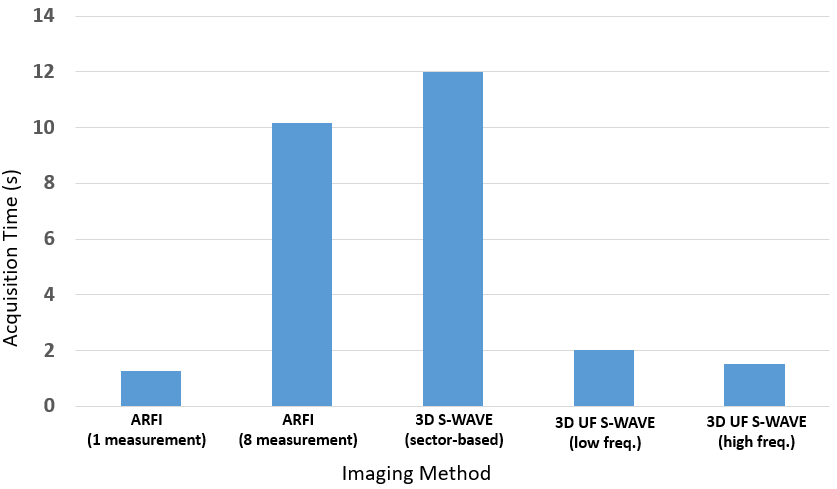}}
		\caption{Acquisition time for different imaging methods.}
		\label{fig:AcqTime}
	\end{center}
\end{figure}

{\subsection{Data Acquisition Time}
	\label{AcqTime}
	Figure~\ref{fig:AcqTime} shows the data collection time for different methods. The ARFI imaging sequence for a single measurement takes 1.2~s, however, the implementation of the method provided for the Verasonics machine uses \textcolor{black}{between 6 to 12} measurements \textcolor{black}{(8 in this work)}. Therefore, the overall data collection using 8 measurements takes 10.16 s assuming no waste{d} time between the measurements. The original 3D S-WAVE method~\cite{zeng2020three} takes 12 s for each patient where the focused beam and sector-based imaging is used. The Ultrafast S-WAVE proposed in this work takes 2 s and 1.5 s for {\textit{deeper} and \textit{shallower}} imaging sequences, respectively. Using plane waves instead of focused beam{s} reduces the number of transducer transmissions, thereby improving the imaging time by 83\% and 88\% compared to the original 3D S-WAVE for \textcolor{black}{the} two proposed sequences.}

%%%%%%%%%%%%%%%%%%%%%%%%% Fig  : data acq time

\section{Discussion}
\label{discussion}

The proposed Ultrafast S-WAVE method provides absolute elasticity of the tissue at ultrafast frame rate and images fast and deep within the body in only 1.5-2 s. The importance of the proposed technique is twofold: First, the low frequency excitation provides deep penetration within the body to examine suspected sites located at higher depths. Second, it decreases the patient examination time while providing 3D volume of tissue elasticity. In fact, one of the disadvantages of the methods generating 3D elasticity images compared to the methods making point measurements is the long acquisition times which increases the role of other undesired motions such as operator hand motion, or motion due to patient coughing, or breathing. Therefore, the displacement estimation could be corrupted leading to an inaccurate elasticity map. However, in our proposed method, acquiring \textcolor{black}{a} 3D ultrasound volume takes 2~s, thereby potentially reducing the motion artifacts in handheld ultrasound examinations. \textcolor{black}{The shear-wave displacement volumes are converted to elasticity maps through solving an inverse problem. As waves propagate in three directions, 3D inversion algorithms provide a more complete analysis of the full wave motion compared to 2D ones where only two components within a slice of the 3D wave propagation are analyzed~\cite{tang2015ultrasoundP1}.} 

%NAFLD ranges from 57\% of overweight individuals attending out-patient clinics to 98\% of nondiabetic obese patients.
%The prevalence of steatosis associated with obesity is dramatically affecting developed countries 

\textcolor{black}{Increasing the accuracy of the elasticity estimation will improve the staging of the liver fibrosis~\cite{petitclerc2017liver}. Liver fibrosis can be caused by the inflammation and liver damage of NAFLD, which is often associated with obesity. The prevalence of obesity is dramatically affecting developed countries. For example, approximately one-third of the United States population is considered obese, with the prevalence of NAFLD in the United States population being approximately 30\%~\cite{vernon2011systematic}. As a result, the need to image deeply is necessary for a large portion of the population. However, the maximum imaging depth for ARFI methods is mostly limited to 8 cm~\cite{frulio2013ultrasound}. Therefore, ARFI is not well suited as a screening tool that must work for all patients. On the other hand, MRE is slow, less accessible, and more prone to motion artifacts compared to ultrasound elastography~\cite{petitclerc2017liver}. So, the Ultrafast S-WAVE has advantages over both ARFI and MRE.}

The Vantage 256 system provides excellent control on tuning different imaging parameters, however, it does not have {regulatory} approval~\cite{food2008information} to be used for the \textit{in vivo} experiments on human subjects. Although our experiment on real tissue is limited to the \textit{ex vivo} bovine liver, the results have shown the potential of the ultrafast elastography to be integrated with the 3D S-WAVE system. A future study {will explore} \textit{in vivo} liver {images} to stage liver fibrosis in patients using Ultrafast S-WAVE {on a system with regulatory approval}.

In this work, the ultrasound frame rate is set to 3000 frames/s which leads to a \textcolor{black}{adequate} number of data samples during each excitation cycle. Furthermore, it is selected within common ranges utilized in the other liver studies~\cite{nightingale2015derivation,budelli2016diffraction,gesnik2020vivo}. Using high frame rate imaging improves the acquisition time substantially for 3D S-WAVE imaging which currently takes 12 s for each subject~\cite{zeng2020three}. It also eliminates {potential} sector artifacts. {Volumetric ultrasound data will also be acquired using electronically steered 2D matrix array transducers in the future to further improve the total exam time.}

\textcolor{black}{It has been shown that the only approach that can theoretically remove longitudinal wave effects is curl filtering~\cite{baghani2009theoretical}. The curl operation involves spatial derivatives of the phasor displacement field and derivatives are affected by noisy displacement estimations. However, the GLUE3D motion estimation algorithm used in this work is robust to noise and therefore provides smooth displacements with high signal-to-noise ratio as shown in~\cite{hashemi2018assessment,hashemi20203d,hashemi2017global}. The reason lies within the simultaneous displacement estimation of several samples and exploitation of the displacement continuity prior throughout the image~\cite{hashemi2017global}. As an alternative approach, one can use spatial high-pass filters to reduce the effects of the compressional waves given the difference in propagation speed of compressional and shear waves. However, the compression artifacts will still be coupled to the shear waves and adversely affect the shear wave speed estimation especially using time-of-flight or phase gradient techniques~\cite{palmeri2008quantifying}.}

%\textcolor{black}{change it: copied from : $https://ieeexplore.ieee.org/abstract/document/7446314?casa_token=QHo0Qt_n6ioAAAAA:VRuddIY8ddiEM_TmXqq6EYrTdGYTKFmYjY3Xbz95nWoVEYDmOWwL1Ta7D_OmA1Wuxvk30dN3eA$}{However the compression artifact is more difficult to decouple from the shear waves, and may cause complications when calculating the shear wave speed using time-of-flight (TOF) or phase gradient methods if not properly suppressed [26], [27].  }

There is a\textcolor{black}{n inherent} trade-off between increasing the frame rate and depth of imaging. The total acquisition time can be improved by increasing the frame rate, however, this reduces the time between two transmit events. Therefore, higher depth imaging will be challenging as longer time is needed for the ultrasound waves to reach the imaging depth and back between two transmit events. For example, \textcolor{black}{in medical applications with the speed of sound of 1540 m/s, single plane wave} imaging at the depth of more than \textcolor{black}{13 cm is} not possible with much faster frame rates such as 6000 frames/s even without considering the system delay. \textcolor{black}{Higher frame rates are possible with lower depth of imaging or higher speed of sound in other materials such as steel where the longitudinal speed of sound is about 5900 m/s.} \textcolor{black}{Furthermore, there is a system trade-off between the frame rate and depth of imaging in Verasonics ultrasound machine. The received ultrasound signals are digitized and stored in local memory prior to transfer to the host computer. So, after each frame collection (including all steered-angle acquisitions), there is a "Transfer To Host" command to transfer the collected data to the computer. For higher depth and more steering angles, the size of data is larger and can takes longer time to be transferred to the host computer which affects the adjusted frame rate in the code. One possible solution can be modifying the default imaging sequence to keep more frames on the local memory and transfer them once to the host computer.}

Some previous studies used sector-based imaging where the region of interest is divided into a number of sectors, and each one is acquired separately at a higher frame rate~\cite{baghani2010high}. {Any error in this approach can lead to disjointed motion measurements between sectors.} Plane wave imaging eliminates potential sector artifacts by insonifying the entire field of view at the same time. Furthermore, it potentially allows the study of dynamic elasticity changes over the cardiac cycle. However, there is a trade-off between image quality and frame rate: the spatial resolution of the data is higher in the focused-beam S-WAVE, whereas the Ultrafast S-WAVE is faster ($>10$ times) in data acquisition. Coherent compounding of multiple plane wave images from successive transmissions can be {explored} to further {investigate} the image quality {trade-off with} frame rate~\cite{montaldo2009coherent}.

\textcolor{black}{A limitation of the LFE algorithm used for elasticity reconstruction in this work is the absence of information on the loss modulus and thus on tissue viscosity. As an alternative, one can utilize the obtained phasors from this method in Voigt, Maxwell, and Zener~\cite{fung2013biomechanics,klatt2007noninvasive} rheological models or a combination of the Voigt model with the finite element methods~\cite{honarvar2012sparsity,honarvar2013curl,honarvar2017comparison} to estimate both elasticity and viscosity of the tissue. These viscoelasticity inversion approaches will be explored in future work.}

% \textcolor{black}{For volumetric imaging with a wobbler transducer in this work, the slices within a data volume are acquired separately, and the ones with the same ?? are placed inside a volume. Therefore, the synchronization between imaging sequence, excitation signal, and motor sweeps is needed. In the other words, the imaging sequence should start/stop at the end/begining of the transducer sweep. Furthermore, the period of the excitation signal should be equal to the imaging time + sweep time in order that start of imaging always takes place at the same state??? of ecitation frequency. In this way, no time delay compensation is needed despite the previous studies[?]. However, it can limit the choice of excitation frequencies to some extent if the minimum imaging time is desired: in multi-frequency excitation, the period of the final excitation signal is larger than the period of indivitual signals. For example, for the excitation frequencies of 40,50,60 Hz, the period of the final signal is 100 ms. If the frequencies are modified to 40,50,70 Hz, the period of the signal  }

% 2- M. Honarvar, J. Lobo, O. Mohareri, S.E. Salcudean, R. Rohling
% Direct vibro-elastography FEM inversion in Cartesian and cylindrical coordinate systems without the local homogeneity assumption
% Phys Med Biol, 60 (2015), pp. 3847-3868

\textcolor{black}{3D ultrasound imaging can provide further information as the real tissue is not isotropic, but it can get affected by acoustic shadows that have been shown to be either helpful in detecting lesions and calcifications, or destructive as artifacts in image processing tasks such as 3D reconstruction, segmentation, and image registration. Acoustic shadows appear as low SNR regions in the ultrasound volume due to highly reflecting structures such as bones and ribs which have significantly different impedances compared to soft tissue.  Integration of displacement continuity regularization over time that is used in GLUE3D algorithm can alleviate or ignore the effect of shadowing artifacts. 3D displacement estimation is also preferred compared to 1D and 2D techniques over a volume of data given that biological tissues are nearly incompressible and applying any force on them (which is the case in elastography) leads the tissue to expand in all three directions. So, lateral and elevational displacements can add further information for elasticity reconstruction algorithms and provide more accurate elasticity values~\cite{hashemi20203d}.}

\section{Conclusion}
\label{conclusion}

In this work, we introduced Ultrafast S-WAVE to {advance} 3D S-WAVE with ultrafast {imaging}. The proposed method measures absolute elasticity of the tissue, and was applied to phantom and \textit{ex vivo} data at the frame rate of 3000 frames/s. The advantages of the proposed method include 3D elasticity estimation and lower patient examination time. {The overall data collection time improved by 83-88\% compared to the original 3D S-WAVE.} Furthermore, using the curl of 3D motion field improved the reconstruction results. Future work will examine \textit{in vivo} liver data to stage liver fibrosis in patients using Ultrafast S-WAVE and also reduce unwanted motion artifacts due to the patient breathing or pulsatile motions. Furthermore, plane-wave S-WAVE eliminated the sector artifacts with standard S-WAVE, dynamic elasticity changes with the heart beat can be studied. The dependency of the {\textit{ex vivo}} liver elasticity values {on} frequency was also shown in this work and should be investigated further together with dynamic elasticity changes over the cardiac cycle. %\textcolor{black}{An implementation of the data acquisition and synchronization with external vibration source and wobbler transducer used in this work is publicly available from the Web site of the corresponding author.}    

\section*{Acknowledgments}
This work was supported by the Canadian Institutes of Health Research, and the Natural Sciences and Engineering Research Council of Canada and by the CA Laszlo Chair in Biomedical Engineering held by Professor Salcudean. The authors would like to thank \textcolor{black}{Dr. Kathryn Nightingale, Dr. Mark Palmeri,} Dr. Mohammad Honarvar and Dr. Damien Garcia for valuable discussions.

%\section*{References}
%%Harvard
\bibliographystyle{model2-names.bst}\biboptions{authoryear}
\bibliography{ref}

\end{document}